\documentclass[12pt]{article}
\usepackage{graphicx}% Include figure files
\usepackage{amsmath,amsthm}
\usepackage{amsfonts}
\usepackage{amssymb}
\usepackage{bm}% bold math
\usepackage[square]{natbib}
\def\upi{\pi}
\def\rv{{\mathbf r}}   % radii
\def\kv{{\mathbf k}}  % wavevector
\def\xv{{\mathbf x}}  % coordinate wavevector

\date{}% It is always \today, today,
\title{The  Phillips spectrum \\[5pt]  and a model of  wind-wave dissipation}
\author{Sergei I. Badulin$^{1,2}$ \& Vladimir E. Zakharov$^{2,3}$\\
{\small $^1$P.~P.~Shirshov Institute of Oceanology of the Russian Academy of Sciences} \\
{\small $^2$Skolkovo Institute of Science and Technology,   Moscow, Russia}\\
{\small $^3$University of Arizona, Tuscon, USA}}

\begin{document}
% Force line breaks with \\

\maketitle

 \begin{abstract}
 We consider an extension of the kinetic equation developed by Newell \& Zakharov \citet{NewellZakharov2008}. The new equation takes into account not only the resonant four-wave interactions but also the dissipation associated with the wave breaking. A dissipation function that  depends on the spectral energy flux is introduced into the equation. This function is determined up to a functional parameter, which optimal choice should be made based on comparison with experiment. A kinetic equation with this dissipation function  describes the transition from the Kolmogorov-Zakharov spectrum $ E (\omega) \sim \omega^ {- 4} $ to the Phillips spectrum $ E (\omega) \sim \omega ^ {-5}$  usually observed experimentally. The version of the dissipation function expressed in terms of the energy spectrum can be used for wave modelling and prediction of sea waves.
 \end{abstract}
%%%%%%%%%%%%%%%


\section{Introduction}
It is well known that the spectra of sea waves both in the presence and in the absence of wind have power-like tails. The shape of the tails in the short-wave range is universal and is given by the famous Phillips spectrum \citep{Phillips58}
\begin{equation}
\label{eq1}
E(\omega) = \alpha_{Ph} g^2 \omega^{-5}
\end{equation}
Here $ \alpha_{Ph} = 0.0081 $ is the Phillips constant. Phillips expressed the fair idea that his spectrum owes its existence and robustness to the phenomenon of wave breaking. However, the authentic hypothesis that the wave field in this asymptotic range is an ensemble  of the Stokes limiting waves \citep{Kuznetsov2004} is refuted by the fact that the mean-square steepness of the Stokes limiting waves $ \mu = \langle \nabla \eta^2 \rangle^{1/2} \approx 0.329$ \citep{Lushnikov2017,DyachenkoS2016} ($ \eta $ being the free surface elevation, angle brackets mean  averaging in space) significantly exceeds the steepness of even the most severe waves ($ \mu \simeq 0.1 $) observed in the ocean. In addition, the Stokes waves of high amplitude are heavily unstable.

The correct interpretation of the Phillips spectrum was proposed in the already mentioned work of Newell \& Zakharov \citep{NewellZakharov2008}. It has been shown that the `Phillips Sea' is an ensemble of localized breakers, uniformly distributed through inverse scales. At the same time, Phillips himself noted \citep {Phillips85} that the maximum breaker scale is approximately an order of magnitude smaller than the length of the dominant wave. Analyzing the numerous experiments \citep{Kitai62, Toba1973a,KawaiEtal1977,Mitsuyasu80,Forristall81,Kahma81} Phillips showed that a universal spectrum $ E (\omega) \sim \omega^{- 4} $ is also realized in the intermediate range of scales. Phillips, however, suggested that this spectrum is the result of the simultaneous action of three factors: wind pumping, nonlinear wave interaction, and dissipation. This concept is still quite widespread, however, it is erroneous if only because of the establishment of the $ \omega^{- 4} $ spectrum in the numerical simulation of swell \citep{BadulinZakharov2017swell}. In addition, it is definitely shown now that in the frequency range $ \omega_p <\omega <3.5 \omega_p $ the nonlinear wave-wave interactions is the dominating physical effect \citep{ZakharovBadulin2011DAN, Zakharov2010Scr, ZBGP2019}. For this reason, the theoretical explanation of the spectrum of $ E (\omega) \sim \omega^{-4} $ is quite simple: this is the exact solution to the stationary Hasselmann equation. This fact was established by Zakharov \& Filonenko back in 1966 \citep{ZakhFil66}.

The spectrum in the intermediate region looks as follows
\begin{equation} \label{eq2}
E(\omega,\theta) = 2C_p\frac {P^{1/3} g^{4/3}}{\omega^4}
\end{equation}
Here $ P $ is the energy flux into the region of large wave numbers and $ C_p $ is the Kolmogorov constant. According to the calculations of Geogjaev \& Zakharov \citep{Geogjaev2017} $ C_p \approx 0.203 $. The spectrum (\ref{eq2}) is just a special case of the weakly turbulent Kolmogorov-Zakharov (KZ) spectra described in detail in the monograph \citep{ZakhFalLvov92}.

The remarkable feature of the  Phillips'  dissipation function proposed in \citep{Phillips85} is a physically transparent  meaning of the function. An attempt to improve the Phillips' dissipation function is the starting point of the  article. At the end of the paper we present the versions of the function that we consider to be `work horses'. The selection of an optimal one can be made on the basis of extensive numerical  experiments.

\section{Phillips' spectrum and the asymptotic theory of water waves}\label{sec:2}
%%%%%%%%%%
 The Hasselmann kinetic equation \cite{Hass1962} for a spatial spectrum of wave action $N_\kv$ of wind-driven waves is written in the following form
\begin{equation}
\label{eq2a}
 \frac{\partial N_{\kv}}{\partial t} +
\nabla_{\kv} \omega_{\kv} {\nabla_\rv N_{\kv}} = S_{nl}+S_{in}+S_{diss}.
\end{equation}
Subscripts $\kv,\,\rv$ for  $\nabla$ are used for gradients in wavevector $\kv$  and in coordinate $\rv$  correspondingly. For $N_\kv(\xv,t)$ and $\omega_\kv$ it means dependence on wavevector. The term $S_{nl}$ in (\ref{eq2})  is responsible for four-wave resonant interactions.   Terms $S_{in}$ and $S_{diss}$ represent correspondingly input of wave action from wind and  dissipation. In contrast to the theoretically based term $S_{nl}$ derived from the first principles, the description of $S_{in}$ and $S_{diss}$ is mostly based on phenomenological parameterizations \cite[e.g.][]{WisePaper2007}. It gives very high dispersion of estimates of $S_{in}$ and $S_{diss}$ in wave modeling and forecasting \cite{BPRZ2005,Zakh2018,ZBGP2019}. Validity and physical correctness of the empirically based terms $S_{in}$ and $S_{diss}$ are generally beyond critical consideration: quantitative aspects are dominating over obvious questions on  physical relevance.  In many cases  these assumptions can be validated in comparison against results of direct simulations within the dynamical phase-resolving models \citep[e.g.][]{Dyachenko2004a,AnnShrira2006jfm,KorotkevichEtal2008,KorotkZakh2013,Korotkevich2019lett}.

The collision integral $S_{nl}$
\begin{equation}\label{eq3}
\begin{array}{lll}
    S_{nl}(\kv,\xv,t)&=&\upi g^2 \int |T_{0123}|^2\left(N_0 N_1 N_2+N_1 N_2 N_3-N_0 N_1 N_2 - N_0 N_1 N_3\right)\\[7pt]
    &\times& \delta(\kv + \kv_1 - \kv_2 - \kv_3)\delta(\omega_0+\omega_1-\omega_2 -\omega_3)d \kv_1 d\kv_2 d\kv_3
    \end{array}
\end{equation}
plays a key role in our study. Explicit formulas can be found in many papers \cite[e.g.][]{BPRZ2005}. Here we focus on general properties of the term. For deep water waves the dispersion relation $\omega(\kv)=\sqrt{g |\kv|}$ and kernels $T_{0123}$  are power-law functions of vector $\kv$ that leads to the homogeneity property for the kernels
\begin{equation}\label{eq4}
    |T(\kappa \kv_0,\kappa \kv_1,\kappa \kv_2,\kappa \kv_3)|^2=\kappa^6 |T( \kv_0, \kv_1,\kv_2, \kv_3)|^2,
\end{equation}
and for the collision integral in terms of wave vectors
\begin{equation}\label{eq5}
    S_{nl}[\kappa \kv, \nu N_\kv] = \kappa^{19/2} \nu^3 S_{nl}[ \kv,  N_\kv]
\end{equation}
or as dependence on frequency $\omega$
\begin{equation}\label{eq6}
    S_{nl}[\upsilon \omega, \nu N_\omega] = \upsilon^{11} \nu^3 S_{nl}[ \omega,  N_\omega]
\end{equation}
($\kappa,\,\upsilon,\,\nu$ are arbitrary positive coefficients). The basic assumption of the  approach of weak nonlinearity  is smallness of wave period $T$ as compared with time scale of nonlinear interactions $T_{nl}$ can be written as follows
\begin{equation}\label{eq7}
    \frac{T}{T_{nl}}=\frac{1}{\omega_\kv N_\kv} \frac{dN_\kv}{d t}=\frac{S_{nl}}{\omega_\kv N_\kv} \ll 1.
\end{equation}
It can break at long time and/or for sufficiently short waves. This is not the case of special distributions, the so-called generalized Phillips spectra, when ratios in (\ref{eq7}) do not depend on wave scale, i.e. the assumption of weak nonlinearity inherits the property of the initial wave field \cite{NewellZakharov2008}.  For deep water waves the classic Phillips spectrum being written for wave energy
\begin{equation}\label{eq8}
E_\kv \sim |\kv|^{-4} \quad \textrm{or} \quad E_\omega \sim \omega^{-5}
\end{equation}
or wave action
\begin{equation}
N_\kv \sim |\kv|^{-9/2} \quad \textrm{or} \quad N_\omega \sim \omega^{-6}
\label{eq9}
\end{equation}
satisfies the condition (\ref{eq7}) for any stretching parameters $\kappa$, $\upsilon$ and $\nu$ in (\ref{eq5},\ref{eq6}). In other words, the asymptotic approach appears to be formally valid at any wave scale. Moreover, one can prove that  condition (\ref{eq7}) remains valid for every term $S_{nl}^{(n)}$ that represents the resonant interaction of $n$ waves in the asymptotic series of the collision integral (\ref{eq2a}) \cite{NewellZakharov2008}
\begin{equation}\label{eq10}
    S_{nl}=\sum_{n=4}^{\infty} S_{nl}^{(n)}.
\end{equation}
The generalized Phillips spectrum (\ref{eq8},\ref{eq9}) does not obey the conservative kinetic equation (\ref{eq2}) and, hence, it can be realized only as a balance of an external forcing (dissipation) and wave-wave resonant interactions. In this regard, the solution (\ref{eq8},\ref{eq9}) differs from the classic Kolmogorov-Zakharov solutions for direct and inverse cascading \cite{ZakhFil66,ZakhZasl83a}  \cite[see for notations][]{BPRZ2005}
\begin{eqnarray}
& N^{(1)}(\kv)  =  C_p {P^{1/3}}{g^{-2/3}}{|\kv|^{-4}}; \quad N^{(1)}(\omega,\theta)  =  2 C_p P^{1/3} g^{4/3}\omega^{-5} ; \label{eq11}\\[7pt]
& N^{(2)}(\kv)  =  C_q Q^{1/3} g^{-1/2} |\kv|^{-23/6}; \quad  N^{(2)}(\omega,\theta)  =  2 C_q Q^{1/3} g^{4/3}\omega^{-14/3}.  \label{eq12}
\end{eqnarray}
Here
\begin{equation}\label{eq13}
  Q=\int_0^\omega \int_{-\pi}^{\pi}  S_{nl} d\omega d\theta; \quad P=-\int_0^\omega \int_{-\pi}^{\pi} \omega S_{nl} d\omega d\theta
\end{equation}
are wave action and energy fluxes and $C_q,\,C_p$ --- the corresponding Kolmogorov's constants. The collision integral $S_{nl}$ for  solutions (\ref{eq11}, \ref{eq12}) is plain zero (fluxes are constant) and estimates of the kinetic equation validity (\ref{eq7}) requires special care. Following the simplest (but not trivial) way \cite{ZakharovBadulin2011DAN,Zakharov2010Scr} one can split  the nonlinear transfer  term $S_{nl}$ into two parts: the nonlinear forcing $F_\kv$ and definitely positive term of the  nonlinear damping $\Gamma_\kv N_\kv$ ($\Gamma_\kv$ -- the nonlinear damping rate) as follows
\[
S_{nl}=F_{\kv} - \Gamma_\kv N_\kv.
\]
The relaxation rate $\Gamma_\kv$ gives a physically correct estimate of time scale of nonlinear wave-wave interactions in the kinetic equation (\ref{eq2}). In accordance with (\ref{eq7}) the asymptotic approach  breaks when \cite[see eq.17 in ][]{ZakharovBadulin2011DAN}
\begin{equation}\label{eq14}
\Gamma_\kv \omega \simeq 4\upi g |\kv|^{9} N_{\kv}^2 = \pi \omega^{12} g^{-4} N_{\omega}^2 \simeq 1.
\end{equation}
For the Phillips spectrum (\ref{eq8},\ref{eq9}) the dimensionless rate (\ref{eq14}) is determined by the spectrum magnitude only and does not depend on the wave scale. For the direct cascade KZ solution (\ref{eq11}) the criterium of validity becomes
\begin{equation}\label{eq15}
4\upi C_p^{2} g^{-1/3}P^{2/3} |\kv_{br}|= 4 \pi C_p^2 g^{-4/3} P^{2/3}\omega_{br}^2 \simeq 1.
\end{equation}
It can be expressed in terms of wave scale and wind speed using an empirical parameterization of wind-wave spectra in the form \cite{BadulinGeog2019}
\begin{equation}\label{eq16}
  E(\omega)=\int_{-\pi}^{\pi} E(\omega,\theta) d\theta=\beta g u_* \omega^{-4}
\end{equation}
with $u_*$ being the friction velocity, $g$ -- gravity acceleration and emprical coefficient $\beta\approx 0.13$ \cite{Toba1973a,DonelanHui85,BattjesEtal1987}. It gives an estimate \cite[cf.][]{NewellZakharov2008}
\begin{equation}\label{eq17}
  \omega_{br}\approx 0.9\frac{u_*}{g}.
\end{equation}
 For  wind speed $U_{10}=15$m/s (at standard height $10$ meters above the sea surface) the break occurs for wave length about $20$ cm that is quite close to the conventional range of wind-driven waves. This fact leads us to the idea to relate the balance of wave-wave interactions and nonlinear dissipation with the Phillips spectrum that is formally valid in the whole range of wave scales.

\section{A flux-based model of  the Phillips spectrum}
%%%%%%%%%%%%%%%%
The formal criteria of validity of the weakly nonlinear approach (\ref{eq14},\ref{eq15}) can be satisfied by a dissipation function that absorbs spectral flux. A one-dimensional version of the kinetic equation in the flux form \cite[e.g.][]{PRZ2003}
\begin{equation}\label{eq18}
  \frac{d E(\omega)}{d t}=-\frac{\partial P}{\partial \omega}+ S_{diss}(P,\omega)
\end{equation}
describes a balance of the energy spectral flux divergence (the nonlinear transfer term $S_{nl}$, eq.~\ref{eq13}) and the dissipation function $S_{diss}$ that depends on the flux $P$ and frequency $\omega$ only. Following the dimensional argumentation \begin{equation}\label{eq19}
  S_{diss}=-\Psi(P\omega^3/g^2)\frac{P}{\omega}.
\end{equation}
The  term $P/\omega$ obeys the same homogeneity conditions as the nonlinear transfer term $S_{nl}$ (\ref{eq5}), thus, realizing the general principle ``like cures like''. With the same  homogeneity conditions (\ref{eq5}) the dimensionless argument of $\Psi$ can be related to the Phillips saturation function \cite{Phillips1984} and energy (action) spectrum. For the isotropic case one has
\begin{equation}\label{eq20}
  B(\omega)=\frac{\mu_d^2}{2}=\frac{\omega^6 N(\omega)}{2 g^2}=\frac{\omega^5 E(\omega)}{2 g^2} \sim \left(\frac{P\omega^3}{g^2}\right)^{1/3},
\end{equation}
i.e. $B(\omega)$ is proportional to the squared differential wave steepness $\mu_d$. The corresponding integral
{\begin{equation}\label{eq22}
  s^2 = 2 \int_0^\omega B(\omega) \frac{d\omega}{\omega}
\end{equation}
is known as mean square slope. It grows logarithmically with frequency for the Phillips spectrum.
In the Phillips model \cite{Phillips85}  $B(\omega)$ is used as an indicator of the degree of saturation of the wave field tending to a finite limit for the spectrum $\omega^{-5} $ .  This limit  can be easily assessed \citep[cf.][eq.~7]{HwangWang2001} for (\ref{eq1}) as a model of fully developed sea \cite{KomenHass84}
\begin{equation}\label{eq23}
  \lim_{\omega\to \infty} B(\omega) = \frac{\alpha_{Ph}}{2}\approx 4.05\cdot 10^{-3}.
\end{equation}
A similar effect of saturation  can be found  in explicit form for stationary solutions of (\ref{eq19})   with the power-law dependencies
\begin{equation}
\label{eq24}
\Psi=a \left(\frac{P\omega^3}{g^2}\right)^R.
\end{equation}
The stationary solution of (\ref{eq19}) is not unique. The simplest one corresponds to saturation of the dissipation function in the whole range of wave frequencies
\begin{equation}\label{eq25}
  \Psi=a \left(\frac{P\omega^3}{g^2}\right)^R=3
\end{equation}
for arbitrary parameters $\alpha$ and $R$. The second solution describes a transition from a finite flux $P_0$ at $\omega \to 0$ to vanishingly small one at $\omega \to \infty$ with the same saturation of the dissipation function $\Psi\to3$ at high frequencies
\begin{equation}\label{eq26}
\frac{P}{P_0}=\left(1+\frac{a}{3}\left(\frac{P_0\omega^3}{g^2}\right)^R\right)^{-1/R}.
\end{equation}
Solutions of (\ref{eq25},\ref{eq26}) are shown in fig.\ref{figure1} as functions of dimensionless frequency
\begin{equation}\label{eq27}
  \Omega=\left(\frac{\omega^3 P_0}{g^2}\right)^{1/3}\left(\frac{a}{3}\right)^{1/(3R)}
\end{equation}
The degenerate solution $\Psi=3$  (\ref{eq25})   corresponds to infinitely large energy fluxes when $\omega\to 0$ (see solid lines in figs. \ref{figure1}{\it a,b}). Solutions (\ref{eq26}) for different exponents $R$\, show a transition from finite energy flux at low frequencies to the power-law flux decay at $\omega \to \infty$. The dissipation rate $\Psi(P\omega^3/g^2)$ manifests a step-like behavior  for high exponents $R$ near a characteristic dimensionless frequency $\Omega=1$ (fig. \ref{figure1}{\it c}). The energy flux $P$ in (\ref{eq26}) can be converted into spectral density with (\ref{eq5}), thus, showing a transition from the KZ $\omega^{-4}$ to the Phillips spectrum $\omega^{-5}$ ({fig. \ref{figure1} \it d}). High $R$ makes this transition sharper.
\begin{figure}
\begin{center}
\includegraphics[scale=0.55]{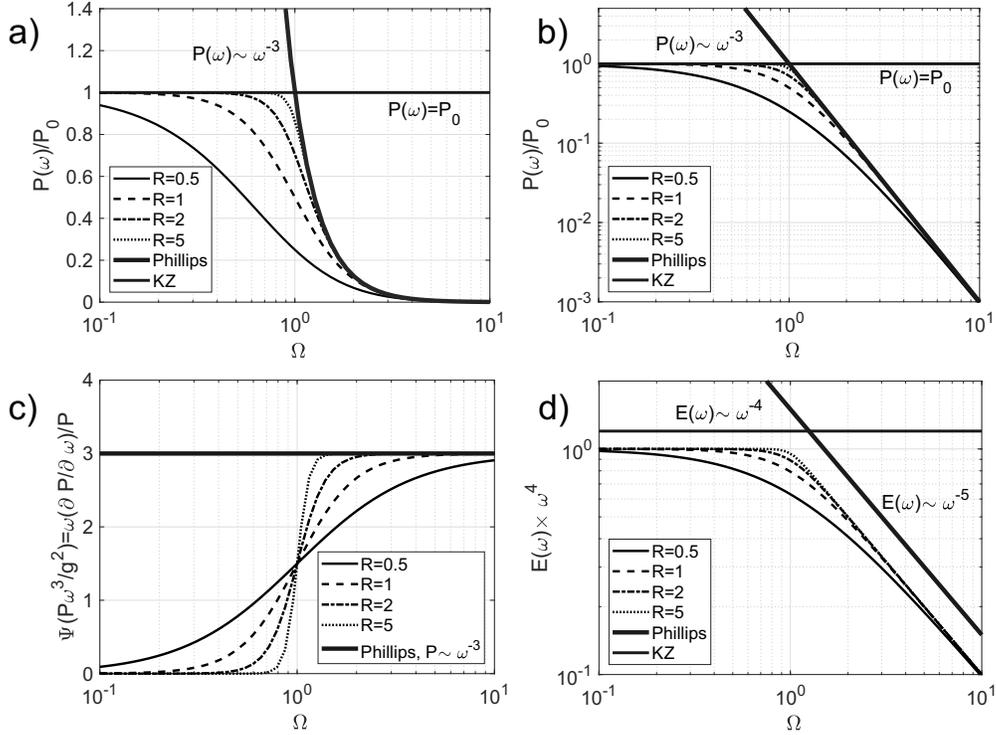}
 \end{center}
  \caption{Stationary solutions for the model (\ref{eq19}). a, b) -- Dimensionless spectral flux for the solutions at different exponents $R$ in semi-log and log-log axes. c) -- dissipation functions at different $R$. $\Psi=3$ for degenerate solution (\ref{eq25}) and one corresponding to power-law dependence (\ref{eq26}) are shown. d) -- compensated spectra derived from homogeneity relationships for spectral fluxes and spectra (\ref{eq5}). The Kolmogorov-Zakharov (\ref{eq11},\ref{eq12}) and Phillips spectra (\ref{eq8}) asymptotes are shown for reference. }
  \label{figure1}
\end{figure}

Solution (\ref{eq26}) allows one for relating the   transition parameters with available experimental data.  Data of Forristall \cite{Forristall81} give for the  transition $\omega_{tr}=g \omega_{tr}/U_{10} \approx 4-5$. For typical inverse wave age of wind driven waves less than $2$ it means the ratio of the transition to peak frequency $\omega_{tr}/\omega_p \approx 2-3$ that agrees well with the observations of P.~Hwang \cite{Hwang2007break}. With experimental parameterization of wave spectra (\ref{eq16}) one has \cite{BadulinGeog2019}
\begin{equation}\label{eq28}
  P_0=0.12\frac{\rho_a}{\rho_w}\frac{u_*^3}{g}
\end{equation}
and an estimate of the unknown coefficient in the dissipation function (\ref{eq24})
\begin{equation}
\label{eq29}
  a=3\left(0.06 \frac{\omega_{cr} u_*}{g}\right)^{-3R}
\end{equation}
The non-zero $R$ means nonlinearity of dissipation in terms of fluxes, while $S_{diss}$ (\ref{eq19}) remains inherently nonlinear as function of spectral density $E(\omega)$ even for $R= 0$.

\section{A local substitute for the dissipation function}
The proposed dissipation function (\ref{eq19}) is nonlocal as one dependent on spectral flux $P$ (\ref{eq13}) and, thus, it is heavy to use it in the wave modeling and forecasting. In this section we show a way to construct `a local substitute' of the dissipation function $S_{diss}$ in the spirit of widely used parameterizations \cite{WisePaper2007}.  Consider power-law distributions in the form
\begin{equation}\label{eq29a}
  N(\kv)=b |\kv|^{-x} \quad {\rm or} \quad E(\omega)=4\pi b\omega^{4-2x}g^{x-2}.
\end{equation}
Energy flux for (\ref{eq29a}) can be calculated analytically \cite{Geogjaev2017}
\begin{equation}\label{eq30}
  P=- \frac{2\pi b^3g^{3x-10}}{12-3x}\omega^{24-6x} F(x)
\end{equation}
where the dimensionless function $F(x)$ depends on the exponent $x$ only.
For the Phillips spectrum $\omega^{-5}$ the exponent $x=9/2$  gives
\begin{equation}\label{eq31}
  \frac{P\omega^3}{g^2}=\frac{ F(9/2)}{48\pi^2}\frac{E\omega^5}{g^2}.
\end{equation}
For the saturated state the dimensionless  energy  dissipation rate $\Psi=3$ in (\ref{eq25}) for the Phillips spectrum with (\ref{eq20}) and $F(9/2)\approx 327$ \cite{Geogjaev2017} becomes
\begin{equation}
\label{eq32}
\gamma_E=\frac{S_{diss}}{\omega E}= \frac{3P}{\omega^2 E}=\frac{F(9/2)}{16\pi^2}\left(\frac{E\omega^5}{g^2}\right)^2 \approx 2.07 \left(\frac{E\omega^5}{g^2}\right)^2.
\end{equation}
Similar estimates  for the experimentally-based dissipation function developed by Mark Donelan \cite{Donelan2001} in terms of the Phillips saturation function \begin{equation}\label{eq33}
S_{diss}=  36\omega E(\kv) \left( B(\kv)\right)^n
\end{equation}
with $n=2.5$ ($R=0.5$ in  our flux presentation) give $4$ times lower values than the theoretical ones (\ref{eq32},\ref{eq20})
\begin{equation}\label{eq33a}
  \gamma_E=1.36 \cdot 10^{-4} \gg \gamma_E^{Donelan}=36 \cdot B^{2.5} (\omega)\approx 3.8\cdot 10^{-5}.
\end{equation}
<<A correction>> (\ref{eq33}) proposed by Donelan that takes, in his opinion, the effect of long waves on the short-wave range \citep[][eq.6]{Donelan2001},
\begin{equation}\label{eq33b}
S_{diss}= 36\omega E(\kv) (1+500\cdot s^2)^2\left( B(\kv)\right)^n
\end{equation}
changes the estimate (\ref{eq33a}) dramatically because of the big multiplier (500 !!!) of formally small value $s$ (\ref{eq22}). The conservative estimate $s=0.02$ \citep[][fig.~1]{HwangWang2001} gives  $\gamma_E^{Donelan}\approx 46\cdot 10^{-4}$ now more than one order higher than the theoretical value (\ref{eq33a}).

The above example demonstrates problems of experimental estimates of the dissipation rates where, even for the same paper  \citep{Donelan2001}, the dispersion of values can exceed two orders of magnitude. At the same time, one should emphasize the qualitative similarity of (\ref{eq33a},\ref{eq33b}) and some operational dependencies \citep[e.g.][]{WesthuysenAl2007} with their  theoretical counterparts developed in this paper. Formulas (\ref{eq33a},\ref{eq33b}) operate exclusively with parameters of wave field, thus, reflecting an inherent physical link of the phenomenon of breaking to the intrinsic wave dynamics. Effects of wind do not enter these dependencies explicitely.

Thus, the simple model of the   dissipation (\ref{eq18}) shows its consistency with the experimental results. The key physical effect of saturation of the dissipation  (\ref{eq25}) taken into account by empirical dependencies \citep{Donelan2001} makes the issue of particular dependence of the function on sea state less important. One can propose an ansatz of the dissipation function that reflects its rather general features:
\begin{itemize}
  \item a characteristic scale (frequency) of transition from the KZ to the Phillips spectrum  associated with the dimensionless frequency $\Omega=1$ in (\ref{eq26},\ref{eq27}) found experimentally at $\omega_{tr}\approx 3-4 \omega_p$;
  \item nonlinear dependence on dimensionless energy spectrum (\ref{eq32}) that is responsible for the saturation effect of the dissipation function and expressed  in terms of dimensionless energy, differential steepness $\mu_d$ (\ref{eq20}) or the Phillips saturation function $B(\omega)$ (\ref{eq33}).
\end{itemize}
The result can be written as follows
\begin{equation}\label{eq35}
  S_{diss}(\omega)=C_{Phillips} \omega \mu^4 E(\omega) \Theta(\omega-\omega_{tr})
\end{equation}
 where $\Theta$ is the Heaviside function that determines the transition of the Kolmogorov-Zakharov spectrum to the Phillips one. We showed the correspondence of the dissipation function (\ref{eq35}) to the problem of the Phillips spectrum saturation earlier \cite{BZ2012}. An alternative version of the dissipation function has been used recently in \citep{Korotkevich2019lett} where the transition to the Phillips spectrum has been provided by a threshold wave steepness $\mu_d$ value. The choice between these two options of the KZ-to-Phillips transition can be made on the basis of extensive simulations. This is the authors' nearest plan.

\section{Conclusions}
We summarize the paper by brief overview of its key points:
\begin{itemize}
\item A simple model of wind wave dissipation is proposed. This model realizes the classic Phillips spectrum $\omega^{-5}$ as a balance of nonlinear transfer due to wave-wave resonant interactions and nonlinear dissipation;
\item The stationary solutions for the simple model describe a saturation of the nonlinear dissipation function for an arbitrary dependence of the function on dimensionless spectral flux. These solutions correspond to transition from the Kolmogorov-Zakharov spectrum $\omega^{-4}$ to the dissipative Phillips spectrum $\omega^{-5}$;
  \item Parameters of the transition from the KZ to the Phillips spectrum are consistent with experimental findings quantitatively \cite{Forristall81};
  \item A local (in spectral scales) theoretically consistent substitute of the proposed nonlinear dissipation function is developed. The comparison with the experimental nonlinear parameterization of the dissipation function by Donelan \cite{Donelan2001} shows qualitative  agreement in the form of dependencies. The dissipation function appears to be almost linear in terms of spectral flux and heavily nonlinear (the dependence is stronger than cubic) for wave energy spectrum. The possibility of a quantitative comparison is substantially complicated by the large scatter of experimental estimates.
\end{itemize}

\paragraph{\underline{Conflict of Interest:}}{The authors declare that they have no conflicts of interest.}
%\noindent \textbf{Acknowledgments}\\

\noindent This research was  supported by Russian Science Foundation grant \# 19-72-30028 with the contribution of MIGO GROUP (http://migogroup.ru).

%\bibliographystyle{unsrt}
%\bibliography{bib_bsi_eng}

%

\end{document}